\documentstyle[12pt]{article}
\def\gsim{\mathrel{\mathpalette\fun >}}
\def\fun#1#2{\lower3.6pt\vbox{\baselineskip0pt\lineskip.9pt
\ialign{$\mathsurround=0pt#1\hfil##\hfil$\crcr#2\crcr\sim\crcr}}}

\newcommand{\be}{\begin{eqnarray}}
\newcommand{\ee}{\end{eqnarray}}

\input{epsfig.sty}

\begin{document}

\Huge{\noindent{Istituto\\Nazionale\\Fisica\\Nucleare}}

\vspace{-3.9cm}

\Large{\rightline{Sezione SANIT\`{A}}}
\normalsize{}
\rightline{Istituto Superiore di Sanit\`{a}}
\rightline{Viale Regina Elena 299}
\rightline{I-00161 Roma, Italy}

\vspace{0.65cm}

\rightline{INFN-ISS 97/10}
\rightline{September 1997}

\vspace{2cm}

\begin{center}

\large{\bf HEAVY-QUARK BINDING AND KINETIC ENERGIES\\ IN HEAVY-LIGHT
MESONS AND\\ THE CONSTITUENT QUARK MODEL\footnote{To appear in Physics 
Letters {\bf B}.}}

\vspace{1cm}

\large{Silvano Simula}

\vspace{0.5cm}

\normalsize{Istituto Nazionale di Fisica Nucleare, Sezione Sanit\`{a},
\\ Viale Regina Elena 299, I-00161 Roma, Italy}

\end{center}

\vspace{1cm}

\begin{abstract}

\indent The spin-averaged binding energy and the hyperfine mass
splitting of heavy-light mesons are investigated within the constituent
quark model as a function of the inverse heavy-quark mass. It is shown
that the so-called heavy-quark kinetic energy, $-\lambda_1 / 2m_Q$,
may differ remarkably from the non-relativistic expectation $\langle
p^2 \rangle / 2m_Q$, thanks to relativistic effects in the effective
interquark potential for heavy-light mesons, which may yield
substantial $1 / m_Q$ corrections to the heavy-quark static limit. The
determination of the difference of the hadronic parameter $\lambda_1$
in the $B_{u(d)}$ and $B_c$ mesons can provide information about the
strength of relativistic effects in the interquark interaction.

\end{abstract}

\vspace{1cm}


\newpage

\pagestyle{plain}

\indent The Heavy Quark Effective Theory ($HQET$) \cite{NEU94} is
widely recognized as a powerful tool to investigate the properties of
hadrons containing a heavy quark. This theory is based on an
effective Lagrangian, written in terms of effective fields, which in
the limit of infinite heavy-quark masses ($\mu_Q \to \infty$) exhibits
a spin-flavour symmetry, the so-called Heavy Quark Symmetry ($HQS$).
Such a symmetry is shared by, but not manifest in $QCD$ and it is only
softly broken by terms of order $\Lambda_{QCD} / \mu_Q$ \cite{IW},
where $\Lambda_{QCD}$ is the $QCD$ scale parameter. The $HQS$ allows to
derive several {\em exact} relations among hadronic properties in the
heavy-quark limit, while at finite values of $\mu_Q$ the $HQET$ allows
to represent physical quantities as a systematic expansion in terms of
powers of the inverse heavy-quark mass, $1 / \mu_Q$, i.e. it organizes
the structure of the power corrections to the $HQS$ limit. However,
neither the $HQS$ nor the $HQET$ can help to predict all the
properties of heavy hadrons, because the calculation of the matrix
elements of the relevant operators requires the knowledge of the
non-perturbative structure of heavy hadrons. In this letter, we will
focus on some of the relevant $HQET$ matrix elements, particularly the
hadronic parameters $\bar{\Lambda}$, $\lambda_1$ and $\lambda_2$,
defined as
 \be
   \bar{\Lambda} & = & lim_{\mu_Q \to \infty} ~ \left[ M^{(j)}(\mu_Q)
   ~ - ~ \mu_Q \right] \nonumber \\
   \lambda_1 & = & \langle M_{\infty} | ~ \bar{h}(v) ~ (iD_{\perp})^2 ~
    h(v) ~ | M_{\infty} \rangle \nonumber \\
   \lambda_2 & = & {g \over 2d_j} ~ \langle M_{\infty} | ~ \bar{h}(v)
   ~ \sigma_{\mu \nu} ~ G^{\mu \nu} ~ h(v) ~ | M_{\infty} \rangle
   \label{eq:DEF}
 \ee
where $M^{(j)}(\mu_Q)$ is the mass of a heavy-light meson with spin
$j$ containing a heavy quark with pole mass $\mu_Q$; $h(v)$ is the
heavy-quark field in the $HQET$ with fixed four-velocity $v$; $|
M_{\infty} \rangle$ is the heavy-meson state in the heavy-quark limit
normalized as $\langle M_{\infty} | \bar{h}(v) h(v) | M_{\infty}
\rangle = 1$; $D_{\perp}^{\mu} \equiv D^{\mu} - (v \cdot D) v^{\mu}$,
with $D^{\mu} \equiv \partial^{\mu} - ig A^{\mu}$ being the
gauge-covariant derivative; $\sigma_{\mu \nu} \equiv {i \over 2}
[\gamma_{\mu}, \gamma_{\nu}]$; $G^{\mu \nu} \equiv {i \over g}
[D^{\mu}, D^{\nu}]$ is the gluon field-strength tensor; $d_j$ is a
spin coefficient, which takes the value $3$ and $-1$ in case of
pseudoscalar and vector mesons, respectively.

\indent The parameter $\bar{\Lambda}$, which is also related to the
trace anomaly of $QCD$ \cite{BIGI} ($\bar{\Lambda} = \langle
M_{\infty} | ~ \bar{h}(v)$ ${\beta(\alpha_s) \over 4 \alpha_s} ~
G_{\mu \nu} ~ G^{\mu \nu}$ $h(v) ~ | M_{\infty} \rangle$, with $\beta$
and $\alpha_s$ being the Gell-Mann-Low function and the running strong
coupling constant), represents the heavy-quark binding energy
and measures the mass of the light degrees of freedom in the static
colour source provided by the heavy quark. The quantity $-\lambda_1
/ 2\mu_Q$ is often referred to as the heavy-quark kinetic energy,
while $-d_j \lambda_2 / 2\mu_Q$ is the leading contribution of the
chromomagnetic interaction to the heavy-meson mass. Thus, the $HQET$
expansion of the meson mass $M^{(j)}(\mu_Q)$ reads as \cite{NEU94}
 \be
    M^{(j)}(\mu_Q) = \mu_Q + \bar{\Lambda} - {\lambda_1 + d_j \lambda_2
    \over 2\mu_Q} + {\rho_1 + d_j \rho_2 \over 4 \mu_Q^2} ~ + ~ O({1
    \over \mu_Q^3})
    \label{eq:HQET}
 \ee
where the parameters $\rho_1$ and $\rho_2$ govern the $1 / \mu_Q^2$
correction terms. All the $HQET$ parameters appearing in Eq.
(\ref{eq:HQET}) contribute to the non-perturbative power corrections
(up to the order $(\Lambda_{QCD} / \mu_Q)^3$) to the free-quark limit
for the inclusive lepton spectrum in heavy-meson semileptonic decays.

\indent Experimental information on the $HQET$ parameter $\lambda_2$ is
directly available from meson spectroscopy through the $B^* - B$ mass
splitting, yielding $\lambda_2 = (M_{B*}^2 - M_B^2) / 4 + O(1 / \mu_Q)
\simeq 0.12 ~ GeV^2$. As for $\lambda_1$ and $\bar{\Lambda}$, several
attempts have recently been performed in order to extract them from
model-independent analyses of the inclusive lepton spectrum of beauty
and charm decays \cite{GREMM,FALK,CHER}. The most recent analysis of
Ref. \cite{GREMM} has provided the values $\bar{\Lambda} = (0.33 \pm
0.11) ~ GeV$ and  $-\lambda_1 = (0.17 \pm 0.10) ~ GeV^2$, where the
quoted uncertainties arise mainly from the non-perturbative
$(\Lambda_{QCD} / \mu_b)^3$ corrections in the inclusive $B$-meson
lepton spectrum.

\indent As far as the theoretical point of view is concerned, the
$HQET$ hadronic parameters have recently been investigated within
the lattice formulation of $HQET$ \cite{MART} and the $QCD$ sum rule
($QCD-SR$) techniques \cite{BIGI,BRAUN,NEU96} (for earlier $QCD-SR$
analyses see Ref. \cite{QCD_SR}). The most interesting  situation
occurs for the $HQET$ parameter $\lambda_1$. In Ref. \cite{BRAUN} the
$QCD-SR$ value $-\lambda_1 = (0.52 \pm 0.12) ~ GeV^2$ has been
obtained, consistently with the lower bound $-\lambda_1 > 3 \lambda_2
\simeq 0.36 ~ GeV^2$ derived in Ref. \cite{BIGI} within the so-called
$QCD-SR$ in the small velocity limit. However, in Ref. \cite{KAP} such
a lower bound has been weakened significantly due to the effects of
high-order radiative corrections. In Ref. \cite{NEU96} a re-analysis
of the $QCD$ sum rules, based on the virial theorem, has yielded a
substantially smaller value, $-\lambda_1 = (0.10 \pm 0.05) ~ GeV^2$.
Moreover, the estimate $-\lambda_1 = -(0.09 \pm 0.14) ~ GeV^2$ has
been extracted from recent lattice $QCD$ simulations \cite{MART}. It
seems clear therefore that a small value of $-\lambda_1$ may be
suggested theoretically and is consistent with available
model-independent information. On the contrary, within the constituent
quark ($CQ$) model the existing predictions for $-\lambda_1$ lie in
the range $0.4 \div 0.6 ~ GeV^2$ (cf. the recent works of Refs.
\cite{FAZIO,KIM}). As a matter of fact, the quark model expectation is
$-\lambda_1 = \langle p^2 \rangle \equiv \langle M_{\infty} | p^2 |
M_{\infty} \rangle$, where $\vec{p}$ is the heavy-quark three-momentum
in the heavy meson and $p^2 = |\vec{p}|^2$. Typically, one has
$\langle p^2 \rangle \simeq 0.4 \div 0.6 ~ GeV^2$, depending on the
detailed shape of the heavy-quark momentum distribution inside the
heavy-light meson (cf. Refs. \cite{FAZIO,KIM}). However, we point out
that the result $-\lambda_1 = \langle p^2 \rangle$ is characteristic
of $CQ$ models in which a non-relativistic interquark potential, which
is more appropriate in case of heavy-heavy mesons, is adopted for the
description of heavy-light mesons, as it has been done in Refs.
\cite{FAZIO,KIM} and \cite{ISGW2}. The aim of this letter is indeed to
show that in case of heavy-light mesons the $CQ$ model ($CQM$)
prediction for $-\lambda_1$ is sharply sensitive to relativistic
effects in the $CQ$ interaction, which may lead to substantial
deviations from the non-relativistic expectation $-\lambda_1 = \langle
p^2 \rangle$.

\indent Let us consider a $Q \bar{q}$ meson, containing a heavy quark
$Q$ with constituent mass $m_Q$ and a lighter partner (anti)quark
$\bar{q}$ with constituent mass $m_{\bar{q}}$. Limiting ourselves to
$S$-wave mesons for simplicity, the effective $CQ$ Hamiltonian can
generally be written in the form (cf. \cite{GI85})
 \be
    \left\{ \sqrt{m_Q^2 + p^2} + \sqrt{m_{\bar{q}}^2 + p^2} + V_{Q
    \bar{q}} \right \} ~ w_{Q \bar{q}}(p^2) |j \mu \rangle = M_{Q
    \bar{q}}^{(j)} ~ w_{Q \bar{q}}(p^2) |j \mu \rangle
    \label{eq:HAM}
 \ee
where $M_{Q \bar{q}}^{(j)}$ is the meson mass, $V_{Q \bar{q}}$ the
effective interquark potential, $w_{Q \bar{q}}(p^2)$ the $S$-wave
radial function and $|j \mu \rangle$ the $CQ$ spin wave
function\footnote{Equation (\ref{eq:HAM}) is not manifestly
covariant; however, following Ref. \cite{SIM96}, one can apply a
unitary transformation to the canonical wave function $w_{Q
\bar{q}}(p^2) |j \mu \rangle$ in order to satisfy explicitly the
requirements of Poincar\'e covariance. Of course, such a unitary
transformation does not affect the eigenvalues of Eq. (\ref{eq:HAM}),
which are the quantities of interest in this work.}.

\indent Assuming that the heavy-quark static limit of $V_{Q
\bar{q}}$ is well defined, $V_{Q \bar{q}}^{\infty} \equiv lim_{m_Q
\to \infty} V_{Q \bar{q}}$, one can construct a heavy-quark expansion
for the eigenvalue $M_{Q \bar{q}}^{(j)}$ of Eq. (\ref{eq:HAM}) which is
similar to the $HQET$ expansion (\ref{eq:HQET}), but expressed in
terms of powers of the inverse heavy-quark constituent mass, namely
 \be
    M_{Q \bar{q}}^j(m_Q) = m_Q + \bar{\Lambda}^{(CQM)} -
    {\lambda_1^{(CQM)} + d_j \lambda_2^{(CQM)} \over 2m_Q} +
    {\rho_1^{(CQM)} + d_j \rho_2^{(CQM)} \over 4 m_Q^2} ~ + ~ O({1
    \over m_Q^3})
    \label{eq:MASS}
 \ee
where
 \be
    \bar{\Lambda}^{(CQM)} & = & \langle M_{\infty}| ~
    \sqrt{m_{\bar{q}}^2 + p^2} + V_{Q \bar{q}}^{\infty} ~
    |M_{\infty}\rangle
    \label{eq:lambda}
 \ee
while $\lambda_1^{(CQM)}$ and $\rho_1^{(CQM)}$ are the relevant $CQM$
parameters in the heavy-quark expansion of the spin-averaged binding
energy $\bar{\Lambda}_{Q \bar{q}}$
 \be
    \bar{\Lambda}_{Q \bar{q}} \equiv {M_{Q \bar{q}}^{(0)} + 3 M_{Q
    \bar{q}}^{(1)} \over 4} - m_Q = \bar{\Lambda}^{(CQM)} - 
    {\lambda_1^{(CQM)} \over 2m_Q} + {\rho_1^{(CQM)} \over 4
     m_Q^2} ~ + ~ O(1/m_Q^3)
    \label{eq:binding}
 \ee
and $\lambda_2^{(CQM)}$ and $\rho_2^{(CQM)}$ appear in the
heavy-quark expansion of the meson mass splitting
 \be
   M_{Q \bar{q}}^{(1)} - M_{Q \bar{q}}^{(0)} = {2 \lambda_2^{(CQM)}
   \over m_Q} - {\rho_2^{(CQM)} \over m_Q^2} ~ + ~ O(1/m_Q^3)
   \label{eq:splitting}
 \ee
If the spin-independent part of the interquark interaction $V_{Q
\bar{q}}$ does not depend on the constituent mass $m_Q$, then one gets
$-\lambda_1^{(CQM)} = \langle p^2 \rangle$, while the implicit
$m_Q$-dependence of the radial wave function $w_{Q \bar{q}}(p^2)$ leads
to $\rho_1^{(CQM)} \neq 0$.

\indent In heavy $Q \bar{Q}$ quarkonia the non-relativistic
approximation is expected to be quite reasonable and, therefore, for
such systems a local potential of the Cornell type, containing an
approximately linear-confining term plus the one-gluon-exchange
contribution, can be adopted. As a matter of fact, the
non-relativistic expansion of the interquark interaction has been
found to be almost adequate for the description of charmonium and
bottonium mass spectra (cf. Ref. \cite{LSG91} and see also Ref.
\cite{BV97} for a recent discussion of semi-relativistic interactions
in heavy quarkonia). In Cornell-type potentials the spin-independent
Coulomb-like term, originating from the one-gluon-exchange
interaction, dominates at small interquark distances and governs the
high-momentum tail of the ground-state meson wave function, affecting
therefore significantly the value of $\langle p^2 \rangle$ (cf. Ref.
\cite{SIM96}).

\indent If a local Cornell-type interaction is adopted in heavy-light
mesons, one typically gets $\langle p^2 \rangle \simeq 0.4 \div 0.6 ~
GeV^2$ (cf. Refs. \cite{FAZIO,KIM}), depending mainly on the specific
value adopted for the (effective) running coupling constant, which
governs directly the strength of the Coulomb-like interaction
term\footnote{In the $ISGW$ model \cite{ISGW2} a local Cornell-type
potential is adopted and the ground-state meson wave function is
approximated by a gaussian ans\"atz, which yields $\langle p^2 \rangle
\simeq 0.28 ~ GeV^2$. However, such an approximation underestimates
significantly the value of $\langle p^2 \rangle$ corresponding to the
{\em exact} ground-state of the $ISGW$ Hamiltonian, because the
effects of the Coulomb-like interaction on the high-momentum tail of
the ground-state meson wave function are not properly taken into
account by a soft gaussian ans\"atz (see Ref. \cite{SIM96}).}.
However, the relativistic effects in the interquark potential are
expected to be significantly larger in heavy-light mesons than in
heavy-heavy ones, mainly because the momentum of the light quark
should be of the order of $\Lambda_{QCD}$ or even larger, which
implies $p / m_{\bar{q}} \gsim 1$. As a consequence, the
non-relativistic expansion of the interquark potential is inaccurate
in heavy-light mesons and the non-relativistic limit can almost be
reached only when both $m_Q$ and $m_{\bar{q}}$ are large enough (see
Refs. \cite{GI85,SIM96,BV97}). The main qualitative features of the
relativistic effects in the interquark interaction are the smearing of
the $CQ$ coordinates and the momentum (energy) dependence of the
strength of the interaction (see, e.g., Ref. \cite{GI85}). The former
feature can usually be accomplished via the introduction of a smearing
function, which extends over interquark distances of the order of the
inverse quark masses. The important point is that the relativistic
effects make the various components of the effective interaction $V_{Q
\bar{q}}$ dependent explicitly upon the masses of both interacting
$CQ$'s. In particular, the dependence of $V_{Q \bar{q}}$ upon $m_Q$,
driven by the relativistic corrections, can lead to $1 / m_Q$
corrections to its heavy-quark static limit $V_{Q \bar{q}}^{\infty} =
lim_{m_Q \to \infty} V_{Q \bar{q}}$, yielding $-\lambda_1^{(CQM)} \neq
\langle p^2 \rangle$, while, on the contrary, in case of the
heavy-quark static interaction $V_{Q \bar{q}}^{\infty}$ one always
gets $-\lambda_1^{(CQM)} = \langle p^2 \rangle$, as in the case of the
non-relativistic approximation (i.e., when also $m_{\bar{q}} \to
\infty$). The $1 / m_Q$ corrections to the heavy-quark static
approximation $V_{Q \bar{q}}^{\infty}$ are expected to depend on the
mass $m_{\bar{q}}$ of the partner (anti)quark and, therefore, the
comparison of the meson masses predicted by $V_{Q \bar{q}}$ and its
heavy-quark static limit $V_{Q \bar{q}}^{\infty}$ can provide an
estimate of the relevance of the relativistic effects in the
interquark interaction.

\indent In Ref. \cite{GI85} a phenomenological parametrization of the
smearing function as well as of the energy dependence of the interquark
interaction has been developed and applied to the calculation of the
meson mass spectra. Though the potential model of Godfrey and Isgur
($GI$) \cite{GI85} is far from being fully justified in terms of $QCD$
and, in particular, its relativized features should be considered
only as a first step towards a full relativistic treatment of the
interquark interaction, it is remarkable the fact that the $GI$
potential allows the reproduction of a large number of meson energy
levels from the pion to the upsilon. In what follows we will consider
the relativized $GI$ interaction, $V_{GI}$, and its heavy-quark static
limit, $V_{GI}^{\infty} \equiv lim_{m_Q \to \infty} ~ V_{GI}$.

\indent The spin-averaged binding energy $\bar{\Lambda}_{Q \bar{q}}$
has been calculated by solving Eq. (\ref{eq:HAM}) for the ground-state
of pseudoscalar and vector heavy-light mesons at various values of
the heavy-quark mass $m_Q$. The results obtained at $m_{\bar{q}} =
0.220, ~ 0.419$ and also $1.628 ~ GeV$, which correspond to the $u$-
($d$-), $s$- and $c$-quark constituent masses adopted in Ref.
\cite{GI85}, are reported in Fig. 1 as a function of the inverse
heavy-quark mass, $1 / m_Q$. A quadratic fit in $1 / m_Q$ (see Eq.
(\ref{eq:binding})) has been applied to the calculated values of
$\bar{\Lambda}_{Q \bar{q}}$ for $m_Q \geq 5 ~ GeV$ and the results are
shown in Fig. 1 by the dashed and solid lines. Let us notice that the
same quadratic fits reproduce also the calculated values of
$\bar{\Lambda}_{Q \bar{q}}$ up to $m_Q \simeq 2 ~ GeV$ within few
$\%$. The values obtained for $\bar{\Lambda}^{(CQM)}$,
$\lambda_1^{(CQM)}$ and $\rho_1^{(CQM)}$ are reported in Table 1. It
can clearly be seen that the values of both $\lambda_1^{(CQM)}$ and
$\rho_1^{(CQM)}$, corresponding to the relativized $GI$ interaction
and its static limit $GI^{\infty}$, differ remarkably even when
$m_{\bar{q}}$ is around the $c$-quark mass\footnote{The values of
$\bar{\Lambda}^{(CQM)}$ for the $GI$ and $GI^{\infty}$ potentials
clearly coincide, because the $GI^{\infty}$ interaction is the
heavy-quark static limit of the $GI$ one (see Eq. (\ref{eq:lambda})).}.
Though the precise value of $\lambda_1^{(CQM)}$ and $\rho_1^{(CQM)}$
depends strongly on the way the relativistic effects are parametrized
in the $CQ$ potential, the results presented in  Fig. 1 and Table 1
clearly show that the $CQM$ parameters $\lambda_1^{(CQM)}$ and
$\rho_1^{(CQM)}$ exhibit a large sensitivity to the relativistic
effects in the interquark interaction, which may lead to substantial
deviations from the heavy-quark static expectations. In particular,
the relativistic effects of the $GI$ potential model make
$-\lambda_1^{(CQM)}$ substantially smaller than the heavy-quark static
expectation. Note also that: ~ i) the absolute value of
$\rho_1^{(CQM)}$ increases as the partner (anti)quark mass
$m_{\bar{q}}$ increases; ~ ii) the relativistic effects parametrized
in the $GI$ interaction reduce remarkably the absolute value of
$\rho_1^{(CQM)}$, which, we remind, contributes to the
non-perturbative $(\Lambda_{QCD} / m_Q)^3$ corrections in the inclusive
lepton spectrum for heavy-meson semileptonic decays. Neverthless, the
$GI$ prediction for $\rho_1^{(CQM)}$ at $m_{\bar{q}} = 0.220 ~ GeV$
(see Table 1) is about twice the value of $\rho_1$ ($\simeq 0.03 ~
GeV^3$) quoted in Ref. \cite{GREMM} as the result of the
vacuum-saturation approximation.

\indent In case of the $GI$ potential the values of $\lambda_2^{(CQM)}$
and $\rho_2^{(CQM)}$, obtained  from the quadratic fit of the
calculated meson mass splitting (\ref{eq:splitting}) for $m_Q \geq 5
~ GeV$, are: $\lambda_2^{(CQM)} = 0.149, ~ 0.145$, $0.215 ~ GeV^2$ and
$\rho_2^{(CQM)} = 0.0026, ~ 0.0423$, $0.453 ~ GeV^3$ at $m_{\bar{q}} =
0.220, ~ 0.419$, $1.628 ~ GeV$, respectively. Note that the value of
$\lambda_2^{(CQM)}$ depends only slightly upon the partner (anti)quark
mass $m_{\bar{q}}$, while $\rho_2^{(CQM)}$ turns out to be quite small
at $m_{\bar{q}} = 0.220$ and $0.419 ~ GeV$, but significantly large
for $m_{\bar{q}}$ around the $c$-quark mass. For the $GI^{\infty}$
interaction one clearly has $\lambda_2^{(CQM)} = \rho_2^{(CQM)} = 0$,
because all the hyperfine terms are vanishing in the heavy-quark
static limit.

\indent The values of the $CQM$ parameters $\bar{\Lambda}^{(CQM)}$,
$\lambda_1^{(CQM)}$ and $\lambda_2^{(CQM)}$, calculated at
$m_{\bar{q}} = 0.220 ~ GeV$ using the $GI$ potential, are directly
compared with the $HQET$ hadronic parameters $\bar{\Lambda}$,
$\lambda_1$ and $\lambda_2$ obtained within various approaches. It can
be seen that the $GI$ value for $\bar{\Lambda}^{(CQM)}$ is in agreement
with the result of Ref. \cite{GREMM}, while the $GI$ value of
$\lambda_1^{(CQM)}$ strongly differs (even in sign) with the findings
of Ref. \cite{GREMM}, but it is consistent with the central value of
the lattice $QCD$ simulations of Ref. \cite{MART}. Moreover, the $GI$
value for $\lambda_2^{(CQM)}$ is quite close to the experimental one,
because the $GI$ interaction is tailored to reproduce the hyperfine
meson mass splittings. However, the direct comparison between the
$HQET$ and $CQM$ hadronic parameters is meaningful only if the $CQ$
mass $m_Q$ is assumed to be equal to the pole heavy-quark mass $\mu_Q$
(see the expansions (\ref{eq:HQET}) and (\ref{eq:MASS})). Since in the
$CQ$ model the effects of the frozen degrees of freedom are expected
to be parametrized also via the $CQ$ masses, one can speculate that
$m_Q \neq \mu_Q$ and consider a heavy-quark expansion of the type
 \be
    m_Q = \mu_Q + \bar{\Lambda}^Q - {\lambda^Q \over 2 \mu_Q} +
    {\rho^Q \over 4 \mu_Q^2} + O(1 / \mu_Q^3)
    \label{eq:mQ}
 \ee
which may lead to
 \be
    \bar{\Lambda} & = & \bar{\Lambda}^{(CQM)} + \bar{\Lambda}^Q
    \nonumber \\
    \lambda_1 & = & \lambda_1^{(CQM)} + \lambda^Q \nonumber \\
    \rho_1 & = & \rho_1^{(CQM)} + \rho^Q + 2 \bar{\Lambda}^Q
    \lambda_1^{(CQM)} \nonumber \\
    \lambda_2 & = & \lambda_2^{(CQM)} \nonumber \\
    \rho_2 & = & \rho_2^{(CQM)} + 2 \bar{\Lambda}^Q \lambda_2^{(CQM)}
    \label{eq:parms}
 \ee
In the heavy-quark limit one could simply expect that any difference
between constituent and current quarks disappear, which means
$lim_{\mu_Q \to \infty} (m_Q - \mu_Q) = 0$, i.e. $\bar{\Lambda}^Q = 0$.
This implies $\bar{\Lambda} = \bar{\Lambda}^{(CQM)}$ (as it is
suggested by the comparison with the result of Ref. \cite{GREMM}) and
$\rho_2 = \rho_2^{(CQM)}$, while one still has $\lambda_1 =
\lambda_1^{(CQM)} + \lambda^Q$ and $\rho_1 = \rho_1^{(CQM)} + \rho^Q$
with unknown values for $\lambda^Q$ and $\rho^Q$\footnote{Using
$\lambda_1 = -0.17 \pm 0.10 ~ GeV^2$ \cite{GREMM} and the $GI$ value
$\lambda_1^{(CQM)} = 0.089 ~ GeV^2$, one gets $\lambda^Q = -0.26 \pm
0.10 ~ GeV^2$, which would be interpreted as the average value of the
current heavy-quark momentum squared in the constituent heavy-quark.}.
In order to get rid of any possible relation between $m_Q$ and
$\mu_Q$, one can consider the difference between the parameter
$\lambda_1$ corresponding to different flavours of the partner
(anti)quark; indeed, from Eq. (\ref{eq:parms}) one easily gets 
 \be
    \lambda_1(\bar{q}') - \lambda_1(\bar{q}) =
    \lambda_1(\bar{q}')^{(CQM)} - \lambda_1(\bar{q})^{(CQM)}
    \label{eq:dlambda_1}
 \ee
The results obtained for Eq. (\ref{eq:dlambda_1}) using the $GI$ and
$GI^{\infty}$ potentials are reported in Table 3. It can be seen that
both the $GI$ and $GI^{\infty}$ predictions for the quantity
$\lambda_1(u) - \lambda_1(s)$ are consistent with the recent lattice
result of Ref. \cite{MART}, while the difference $\lambda_1(u) -
\lambda_1(c)$ (i.e., the difference of the hadronic parameter
$\lambda_1$ in the $B_{u(d)}$ and $B_c$ mesons) is sharply sensitive to
the $1 / m_Q$ corrections to the heavy-quark static limit and,
therefore, its determination could provide information on the
relativistic effects in the interquark interaction.

\indent In conclusion, the spin-averaged binding energy and the
hyperfine mass splitting of heavy-light mesons have been investigated
within the constituent quark model as a function the inverse
heavy-quark mass. It has been shown that the so-called heavy-quark
kinetic energy, $-\lambda_1^{(CQM)} / 2m_Q$, may differ remarkably
from the non-relativistic expectation $\langle p^2 \rangle / 2m_Q$,
thanks to relativistic effects in the effective interquark potential,
which may produce substantial $1 / m_Q$ corrections to the heavy-quark
static limit. These corrections, which depend on the mass $m_{\bar{q}}$
of the partner (anti)quark in the heavy meson, have been found to be
significant also for $m_{\bar{q}}$ around the $c$-quark mass.
Moreover, the $1 / m_Q^2$ terms in the heavy-quark expansion of the
meson mass are predicted to be quite small for $m_{\bar{q}}$
around the $u$- and $s$-quark constituent masses, but significantly
larger for $m_{\bar{q}}$ around the $c$-quark mass. Finally, it has
been suggested that the determination of the difference of the hadronic
parameter $\lambda_1$ in the $B_{u(d)}$ and $B_c$ mesons can provide
information about the strength of the relativistic effects in the
interquark interaction.

\vspace{0.5cm}

\noindent {\bf Acknowledgments.} The author thanks Fabio Cardarelli
for helpful discussions and a careful reading of the manuscript.

\newpage

\begin{figure}[htb]

\vspace{-3cm}

\epsfxsize=19cm \epsfig{file=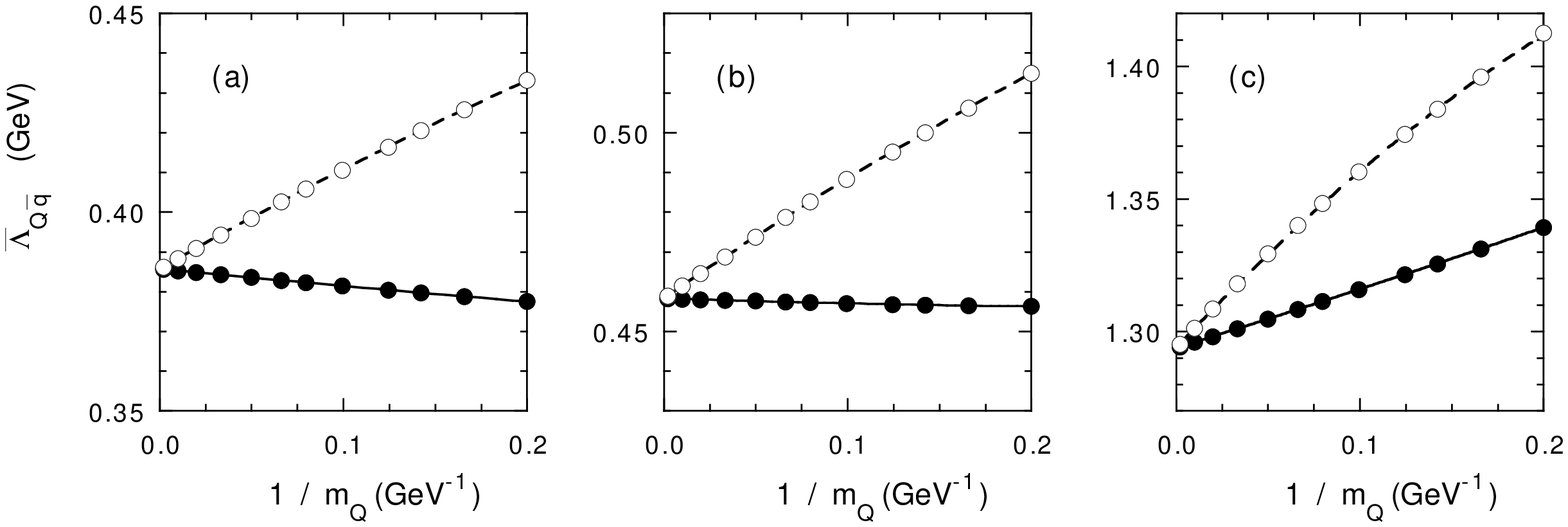}

\vspace{-4cm}

\noindent {\bf Figure 1.} Values of the spin-averaged binding energy
$\bar{\Lambda}_{Q \bar{q}}$ (Eq. (\ref{eq:binding})) as a function of
the inverse heavy-quark constituent mass $1 / m_Q$, obtained from the
ground-state eigenvalue of Eq. (\ref{eq:HAM}) calculated in case of
pseudoscalar and vector heavy-light mesons adopting the $GI$ potential
model \cite{GI85} (full dots) and its heavy-quark static limit
$GI^{\infty}$ (open dots). In (a, b, c) the mass of the partner
(anti)quark is $m_{\bar{q}} = 0.220, 0.419$ and $1.628 ~ GeV$,
respectively. The dashed and solid lines are quadratic fits in $1 /
m_Q$ of the calculated values of $\bar{\Lambda}_{Q \bar{q}}$ for $m_Q
\geq 5 ~ GeV$.

\end{figure}

\vspace{3cm}

\begin{table}[htb]

\noindent {\bf Table 1.} Values of the $CQM$ parameters
$\bar{\Lambda}^{(CQM)}$, $\lambda_1^{(CQM)}$ and $\rho_1^{(CQM)}$,
obtained from the quadratic fit of the calculated values of the
spin-averaged binding energy $\bar{\Lambda}_{Q \bar{q}}$
(\ref{eq:binding}) (see text and Fig. 1). The values of the partner
(anti)quark mass $m_{\bar{q}}$ are given in $GeV$ and correspond to
the $u$- ($d$-), $s$- and $c$-quark constituent masses of Ref.
\cite{GI85}.

\begin{center}

\begin{tabular}{||c|c||c|c|c||} \hline
 $Pot.$ & $CQM ~ param.$ & $m_{\bar{q}} = 0.220$ & $m_{\bar{q}} =
 0.419$ & $m_{\bar{q}} = 1.628$ \\ \hline \hline
               & $\bar{\Lambda}^{(CQM)} ~ (GeV)$ & $~0.386$ & $~0.458$
 & $~1.294$ \\ \cline{2-5}
 $GI$          & $-\lambda_1^{(CQM)} ~ (GeV^2)$  & $-0.089$ & $-0.031$
 & $~0.426$ \\ \cline{2-5}
               & $~\rho_1^{(CQM)} ~ (GeV^3)$     & $~0.063$ & $~0.103$
 & $~0.346$ \\ \hline \hline
               & $\bar{\Lambda}^{(CQM)} ~ (GeV)$ & $~0.386$ & $~0.458$
 & $~1.294$ \\ \cline{2-5}
 $GI^{\infty}$ & $-\lambda_1^{(CQM)} ~ (GeV^2)$  & $~0.523$ & $~0.635$
 & $~1.474$ \\ \cline{2-5}
               & $~\rho_1^{(CQM)} ~ (GeV^3)$     & $-0.514$ & $-0.717$
 & $-2.966$ \\ \hline \hline
\end{tabular}

\end{center}

\end{table}

\newpage

\noindent {\bf Table 2.} Comparison of the values of the $HQET$ hadronic
parameters $\bar{\Lambda}$, $\lambda_1$ and $\lambda_2$ (Eq.
(\ref{eq:DEF})) obtained within various approaches, with the $CQM$
parameters $\bar{\Lambda}^{(CQM)}$, $\lambda_1^{(CQM)}$ and
$\lambda_2^{(CQM)}$ (Eqs. (\ref{eq:lambda}-\ref{eq:splitting})),
calculated in the present work at $m_{\bar{q}} = 0.220 ~ GeV$ using
the $GI$ interaction \cite{GI85}.

\begin{table}[htb]

\begin{center}

\begin{tabular}{||c|c||c|c|c||} \hline
 $Ref.$ & $Method$ & $\bar{\Lambda} ~ (GeV)$ & $-\lambda_1 ~ (GeV^2)$ &
 $\lambda_2 ~ (GeV^2)$ \\ \hline \hline
 $\cite{BIGI}$  & $QCD-SR$     & $> 0.5$         & $>0.36$          &
 $--$            \\ \hline
 $\cite{BRAUN}$ & $QCD-SR$     & $0.4 \div 0.5$  & $0.52 \pm 0.12$  &
 $0.11 \pm 0.04$ \\ \hline
 $\cite{NEU96}$ & $QCD-SR$     & $--$            & $0.10 \pm 0.05$  &
 $0.15 \pm 0.03$ \\ \hline \hline
 $\cite{MART}$  & $lattice$    & $0.18 \pm 0.03$ & $-0.09 \pm 0.14$ &
 $0.07 \pm 0.01$ \\ \hline \hline
 $\cite{GREMM}$ & $exp.$       & $0.33 \pm 0.11$ & $0.17 \pm 0.10$  &
 $\simeq 0.12$   \\ \hline
 $\cite{FALK}$  & $exp.$       & $0.45$          & $0.10$           &
 $\simeq 0.12$   \\ \hline
 $\cite{CHER}$  & $exp.$       & $--$            & $0.14$           &
 $\simeq 0.12$   \\ \hline \hline
 $\cite{FAZIO}$ & $CQM$        & $0.35$          & $0.66$           &
 $--$            \\ \hline
 $\cite{KIM}$   & $CQM$        & $0.5 \div 0.6$  & $0.4 \div 0.6$   &
 $--$            \\ \hline \hline
 $this ~ work$  & $rel. ~ CQM$ & $0.386$         & $-0.089$         &
 $0.149$         \\ \hline \hline
\end{tabular}

\end{center}

\end{table}

\vspace{3cm}

\noindent {\bf Table 3.} Values of the difference of the $HQET$
parameters $\lambda_1$ (in $GeV^2$) corresponding to different
flavours of the partner (anti)quark predicted by the $GI$ \cite{GI85}
and $GI^{\infty}$ potentials through Eq. (\ref{eq:dlambda_1}). The
result of the lattice $QCD$ simulation of Ref. \cite{MART} is also
reported.

\begin{table}[htb]

\begin{center}

\begin{tabular}{||c||c|c||} \hline
 $Potential$ & $\lambda_1(u) - \lambda_1(s)$ & $\lambda_1(u) -
 \lambda_1(c)$ \\ \hline \hline
 $GI$                    & $0.058$ & $0.515$ \\ \hline
 $GI^{\infty}$           & $0.112$ & $0.951$ \\ \hline \hline
 $lattice ~ \cite{MART}$ & $0.09 \pm 0.04$ & $--$ \\ \hline \hline
\end{tabular}

\end{center}

\end{table}

\end{document}